\documentclass[12pt]{article}
\usepackage{amssymb}
\usepackage{amsfonts}
\usepackage{myart}

\input tcilatex
\begin{document}

\title{Discrete space-time geometry and skeleton conception of particle
dynamics}
\author{Yuri A.Rylov}
\date{Institute for Problems in Mechanics, Russian Academy of Sciences,\\
101-1, Vernadskii Ave., Moscow, 119526, Russia.\\
e-mail: rylov@ipmnet.ru\\
Web site: {$http://rsfq1.physics.sunysb.edu/\symbol{126}rylov/yrylov.htm$}\\
or mirror Web site: {$http://gasdyn-ipm.ipmnet.ru/\symbol{126}%
rylov/yrylov.htm$}}
\maketitle

\begin{abstract}
It is shown that properties of a discrete space-time geometry distinguish
from properties of the Riemannian space-time geometry. The discrete geometry
is a physical geometry, which is described completely by the world function.
The discrete geometry is nonaxiomatizable and multivariant. The equivalence
relation is intransitive in the discrete geometry. The particles are
described by world chains (broken lines with finite length of links),
because in the discrete space-time geometry there are no infinitesimal
lengths. Motion of particles is stochastic, and statistical description of
them leads to the Schr\"{o}dinger equation, if the elementary length of the
discrete geometry depends on the quantum constant in a proper way.
\end{abstract}


\section{ Introduction}

Is the space-time geometry discrete, or is it continuous, but equipped by
quantum properties? The quantum properties associate with some portions,
quanta and discreteness. However, it seems intuitively, that the space-time
geometry is rather discrete, than it has mystic quantum properties, because
the concept of discreteness does not contain anything mysterious.
Nevertheless if one consider the scales, which are much more, than the
elementary length $\lambda _{0}$ (characteristic length of the discrete
geometry), it is of no importance, whether the space-time geometry is
discrete or continuous.

Contemporary space-time geometry is a differential geometry. All
concepts of the differential geometry are based on the concept of
continuity. Such concepts (manifold, dimension, coordinate system,
differential equations of dynamics) can be introduced and used
only in a continuous space-time geometry. Besides, a discrete
geometry is nonaxiomatizable geometry. It means, that a discrete
geometry cannot be obtained as a logical construction, because the
equivalence relation is intransitive, in general, in a discrete
geometry. In its turn the intransitivity of the equivalence
relation is a corollary of the following circumstance.  For
determination of a vector $\mathbf{P}_{0}\mathbf{P}_{1}$ at the
given point $P_{0}$, which is
equivalent to a given vector $\mathbf{Q}_{0}\mathbf{Q}_{1}$ at the point $%
Q_{0}$, one needs to solve two algebraic equations (condition of vectors
parallelism, condition of equality of their lengths). The number of
coordinates (four) of the point $P_{1}$ is more than the number of
equations. As a result there are many solutions, in general. There are many
vectors $\mathbf{P}_{0}\mathbf{P}_{1}$, $\mathbf{P}_{0}\mathbf{P}%
_{1}^{\prime }$,... at the given point $P_{0}$, which are equivalent to the
vector $\mathbf{Q}_{0}\mathbf{Q}_{1}$ at the point $Q_{0}$, but vectors $%
\mathbf{P}_{0}\mathbf{P}_{1}$,
$\mathbf{P}_{0}\mathbf{P}_{1}^{\prime }$,... are not equivalent
between themselves. Such a situation means intransitivity of the
equivalence relation for vectors. Hence, a discrete geometry
cannot be a logical construction, because in any logical
construction the equivalence relation is transitive. It cannot be
constructed by the Euclidean method, which is used for
construction of the proper Euclidean geometry.

On the other hand, the Euclidean method of a geometry construction
is used longer, than two thousands years, and we do not know any
other method. The only Euclidean method of a geometry construction
is learnt in all schools. As a result we perceive hardly the idea
of alternative method of a geometry construction. We cannot
imagine, how one can construct a discrete space-time geometry, if
it is nonaxiomatizable and one cannot use the conventional
Euclidean method. It seems to be simpler to consider continuous
space-time geometry, equipping it by mystic quantum properties,
which imitate discreteness. This imitation appeared to be very
successful. Unfortunately, this imitation is not complete (as well
as the imitation of thermal phenomena by means of the thermogen),
and we are forced to return to idea of discrete space-time
geometry, if we want to describe physical phenomena in microcosm.

A geometry is a science on geometrical objects, on their shape and on their
disposition in the space or in the space-time. The proper Euclidean geometry
is the first geometry, which has been constructed. Geometrical objects of
the Euclidean geometry are usually constructed as combination of fundamental
blocks (point, segment of straight, angle). Properties of these fundamental
blocks and the rules of their combination at the construction of geometrical
objects were formulated as axioms of some logical construction. The proper
Euclidean geometry was studied in the form of a logical construction in the
last two thousand years. As a result practically all scientists believe,
that any geometry is a logical construction, and that any geometry is to be
constructed as a logical construction. It is supposed, that any geometry can
be deduced from a system of properly chosen axioms, and there are no
geometries which cannot be deduced from a proper chosen axiomatics.

This fact is formulated as follows. Only axiomatizable geometries do exist
and there are no nonaxiomatizable geometries. In other words, a geometry is
identified with a logical construction. The belief in the identity of a
geometry and a logical construction was so large, that one uses the term
"geometry" with respect to disciplines, which have a logical structure of
the Euclidean geometry, but have no relation to the science on geometrical
objects and their shape (for instance, symplectic geometry). A geometry,
identified with a logical construction (axiomatizable geometry) will be
referred to as a mathematical geometry. The proper Euclidean is a
mathematical geometry.

A geometry as a science on geometrical objects, on their shape and on their
disposition is described completely by a distance function $\rho \left(
P,Q\right) $ between any two points $P,Q\in \Omega $, where $\Omega $ is a
point set, where the geometry is given. It is more convenient to use world
function $\sigma =\frac{1}{2}\rho ^{2}$ instead of the distance function $%
\rho $, because the world function $\sigma $ is real in any geometry (even
in the space-time geometry of Minkowski). The geometry, which is described
completely by their world function will be referred to as a physical
geometry, because physicists are not interested in the way of the geometry
construction. They are not interested in the circumstance, whether or not
the physical geometry is axiomatizable.

The proper Euclidean geometry as well as the geometry of Minkowski are
continuous geometries. However, there may exist discrete geometries, where
the distance between any two points of the space-time is larger, than some
elementary length $\lambda _{0}$. If characteristic scale of the problem is
much larger, than the elementary length $\lambda _{0}$, one may set $\lambda
_{0}=0$ and consider a continuous geometry. However, in microcosm, where
characteristic scale is of the order of $\lambda _{0}$, one should consider
a discrete space-time geometry.

At the conventional construction of the Euclidean geometry one uses such
concepts as manifold, dimension, coordinate system, linear vector space,
which might be used only in continuous geometries. Constructing a discrete
geometry as a generalization of the proper Euclidean geometry, one may not
use these concepts. The only concept, which may be used in both continuous
geometry and in the discrete one, is the distance $\rho $. But the distance $%
\rho $ is to be introduced as a fundamental quantity. In the Riemannian
geometry the distance $\rho $ is introduced as an integral along the
geodesic from the infinitesimal distance
\[
ds=\sqrt{g_{ik}dx^{i}dx^{k}}
\]%
Such a method of introduction of the distance $\rho $ is inadequate in the
discrete geometry, because it uses infinitesimal distance, which does not
exist in the discrete geometry. Besides, in the case, when there are several
geodesics, connecting two points, one obtains many-valued expressions for
the distance or for the world function. Many-valued world function is
inadmissible in a geometry.

To construct a discrete geometry, one needs to represent the proper
Euclidean geometry in terms of the distance $\rho $ (or in terms of the
world function $\sigma =\frac{1}{2}\rho ^{2}$) and to use this
representation for generalization of the proper Euclidean geometry $\mathcal{%
G}_{\mathrm{E}}$ on the case of a discrete geometry $\mathcal{G}_{\mathrm{d}%
} $. Representation of a geometry in terms of a world function will be
referred to as $\sigma $-immanent representation. The $\sigma $-immanent
representation of the proper Euclidean geometry $\mathcal{G}_{\mathrm{E}}$
is always possible.

The main geometrical objects and concepts of $\mathcal{G}_{\mathrm{E}}$: (1)
straight segment $\mathcal{T}_{\left[ PQ\right] }$ between the points $P$
and $Q$, (2) scalar product $\left( \mathbf{P}_{0}\mathbf{P}_{1}.\mathbf{Q}%
_{0}\mathbf{Q}_{1}\right) $ of two vectors $\mathbf{P}_{0}\mathbf{P}_{1}$, $%
\mathbf{Q}_{0}\mathbf{Q}_{1}$, (3) linear dependence of $n$ vectors $\mathbf{%
P}_{0}\mathbf{P}_{1},\mathbf{P}_{0}\mathbf{P}_{2},...\mathbf{P}_{0}\mathbf{P}%
_{n}$, (4) equivalence $\left( \mathbf{P}_{0}\mathbf{P}_{1}\text{eqv}\mathbf{%
Q}_{0}\mathbf{Q}_{1}\right) $ of two vectors $\mathbf{P}_{0}\mathbf{P}_{1}$,
$\mathbf{Q}_{0}\mathbf{Q}_{1}$ can be expressed in terms of the world
function $\sigma _{\mathrm{E}}$ of the geometry $\mathcal{G}_{\mathrm{E}}$%
\begin{equation}
\mathcal{T}_{\left[ PQ\right] }=\left\{ R|\rho \left( P,R\right) +\rho
\left( R,Q\right) =\rho \left( P,Q\right) \right\} ,\qquad \rho \left(
P,Q\right) =\sqrt{2\sigma \left( P,Q\right) }  \label{c1.1}
\end{equation}%
\begin{equation}
\left( \mathbf{P}_{0}\mathbf{P}_{1}.\mathbf{Q}_{0}\mathbf{Q}_{1}\right)
=\sigma \left( P_{0},Q_{1}\right) +\sigma \left( P_{1},Q_{0}\right) -\sigma
\left( P_{0},Q_{0}\right) -\sigma \left( P_{1},Q_{1}\right)  \label{c1.2}
\end{equation}%
\begin{equation}
F_{n}\left( \mathcal{P}_{n}\right) \equiv \det \left\vert \left\vert \left(
\mathbf{P}_{0}\mathbf{P}_{i}.\mathbf{P}_{0}\mathbf{P}_{k}\right) \right\vert
\right\vert =0,\qquad i,k=1,2,...n  \label{c1.3}
\end{equation}%
where%
\begin{equation}
\mathcal{P}_{n}=\left\{ P_{0},P_{1},...P_{n}\right\}  \label{c1.4}
\end{equation}%
and the scalar product of two vectors is defined by the formula (\ref{c1.2}%
), which in the case of common origin has the form%
\begin{equation}
\left( \mathbf{P}_{0}\mathbf{P}_{1}.\mathbf{P}_{0}\mathbf{P}_{2}\right)
=\sigma \left( P_{0},P_{2}\right) +\sigma \left( P_{0},P_{1}\right) -\sigma
\left( P_{1},P_{2}\right)  \label{c1.4a}
\end{equation}%
Equivalence $\left( \mathbf{P}_{0}\mathbf{P}_{1}\text{eqv}\mathbf{Q}_{0}%
\mathbf{Q}_{1}\right) $ of two vectors is described by two relations%
\begin{equation}
\left( \mathbf{P}_{0}\mathbf{P}_{1}\text{eqv}\mathbf{Q}_{0}\mathbf{Q}%
_{1}\right) :\quad \left( \mathbf{P}_{0}\mathbf{P}_{1}.\mathbf{Q}_{0}\mathbf{%
Q}_{1}\right) =\left\vert \mathbf{P}_{0}\mathbf{P}_{1}\right\vert \cdot
\left\vert \mathbf{Q}_{0}\mathbf{Q}_{1}\right\vert \wedge \left\vert \mathbf{%
P}_{0}\mathbf{P}_{1}\right\vert =\left\vert \mathbf{Q}_{0}\mathbf{Q}%
_{1}\right\vert  \label{c1.5}
\end{equation}%
\begin{equation}
\left\vert \mathbf{P}_{0}\mathbf{P}_{1}\right\vert =\sqrt{2\sigma \left(
P_{0},P_{1}\right) }  \label{c1.6}
\end{equation}%
where $\sigma =\sigma _{\mathrm{E}}$ is the world function of the proper
Euclidean geometry $\mathcal{G}_{\mathrm{E}}$.

These relations are to use for definition of such quantities as $\mathcal{T}%
_{\left[ PQ\right] }$, $\left( \mathbf{P}_{0}\mathbf{P}_{1}.\mathbf{Q}_{0}%
\mathbf{Q}_{1}\right) $, $\left( \mathbf{P}_{0}\mathbf{P}_{1}\text{eqv}%
\mathbf{Q}_{0}\mathbf{Q}_{1}\right) $ and linear dependence, because these
definitions do not refer to dimension, coordinate system, linear vector
space and other means of description of the proper Euclidean geometry. These
definitions contain only such a fundamental geometrical quantity as the
world function. These definitions are written in the coordinateless
invariant form. As a result these definitions may and must be used as
definitions of any physical geometry, i.e. a geometry described completely
in terms of the world function. Any geometrical object can be described in
terms of the world function. The geometrical object properties are
calculated on the basis of its $\sigma $-immanent representation
(representation in terms of the world function) and of the world function
form.

Generalization of these expressions on the case of the discrete geometry $%
\mathcal{G}_{\mathrm{d}}$ is obtained by means of the replacement $\sigma _{%
\mathrm{E}}\rightarrow \sigma _{\mathrm{d}}$, where $\sigma _{\mathrm{d}}$
is the world function of the discrete geometry $\mathcal{G}_{\mathrm{d}}$.
We are to be ready, that properties of concepts (\ref{c1.1}) - (\ref{c1.5})
in $\mathcal{G}_{\mathrm{d}}$ differ strongly from their properties in $%
\mathcal{G}_{\mathrm{E}}$. However,  we have no alternative to
relations (\ref{c1.1}) - (\ref{c1.5}) for definition of these
geometrical quantities in a discrete geometry
$\mathcal{G}_{\mathrm{d}}$.

\textit{Definition 1.1}\emph{: }The physical geometry $\mathcal{G}=\left\{
\sigma ,\Omega \right\} $ is a point set $\Omega $ with the single-valued
function $\sigma $ \ on it
\begin{equation}
\sigma :\qquad \Omega \times \Omega \rightarrow \mathbb{R},\qquad \sigma
\left( P,P\right) =0,\qquad \sigma \left( P,Q\right) =\sigma \left(
Q,P\right) ,\qquad P,Q\in \Omega  \label{b4.1}
\end{equation}

\textit{Definition 1.2:} Two physical geometries $\mathcal{G}_{1}=\left\{
\sigma _{1},\Omega _{1}\right\} $ \ and $\mathcal{G}_{2}=\left\{ \sigma
_{2},\Omega _{2}\right\} $ are equivalent $\left( \mathcal{G}_{1}\mathrm{eqv}%
\mathcal{G}_{2}\right) $, if the point set $\Omega _{1}\subseteq
\Omega _{2}\wedge \sigma _{1}\left( P,Q\right) =\sigma _{2}\left(
P,Q\right) ,\ \ \forall P,Q\in \Omega _{1}$, or $\Omega
_{2}\subseteq \Omega _{1}\wedge \sigma _{2}\left( P,Q\right)
=\sigma _{1}\left( P,Q\right) ,\ \ \forall P,Q\in \Omega _{2}$

\textit{Remark:}\emph{\ }\ Coincidence of \ point sets $\Omega _{1}$ and $%
\Omega _{2}$ is not necessary for equivalence of geometries $\mathcal{G}_{1}$%
\textrm{\ }and\textrm{\ }$\mathcal{G}_{2}$. \ If one demands coincidence of $%
\Omega _{1}$ and $\Omega _{2}$ in the case equivalence of $\mathcal{G}_{1}$
and \ $\mathcal{G}_{2}$, then an elimination of one point $P$ from the point
set $\Omega _{1}$ turns the geometry $\mathcal{G}_{1}=\left\{ \sigma
_{1},\Omega _{1}\right\} $ into geometry $\mathcal{G}_{2}=\left\{ \sigma
_{1},\Omega _{1}\backslash P\right\} $, which appears to be not equivalent
to the geometry $\mathcal{G}_{1}$. Such a situation seems to be
inadmissible, because a geometry on a part $\omega \subset \Omega _{1}$ of
the point set $\Omega _{1}$ appears to be not equivalent to the geometry on
the whole point set $\Omega _{1}$.

According to definition \ the geometries $\mathcal{G}_{1}\mathcal{=}\left\{
\sigma ,\omega _{1}\right\} $ and $\mathcal{G}_{2}\mathcal{=}\left\{ \sigma
,\omega _{2}\right\} $ on parts\ of $\Omega $, $\omega _{1}\subset \Omega $
and $\omega _{2}\subset \Omega $ \ \ are equivalent $\left( \mathcal{G}_{1}%
\mathrm{eqv}\mathcal{G}\right) ,\left( \mathcal{G}_{2}\mathrm{eqv}\mathcal{G}%
\right) $ to the geometry $\mathcal{G}$, whereas the geometries $\mathcal{G}%
_{1}\mathcal{=}\left\{ \sigma ,\omega _{1}\right\} $ and $\mathcal{G}_{2}%
\mathcal{=}\left\{ \sigma ,\omega _{2}\right\} $ are not equivalent, in
general, if $\omega _{1}\nsubseteqq \omega _{2}$ and $\omega _{2}\nsubseteqq
\omega _{1}$. Thus, the relation of equivalence is intransitive, in general.
The space-time geometry may vary in different regions of the space-time. It
means, that a physical body, described as a geometrical object, may evolve
in such a way, that it appears in regions with different space-time geometry.

The space-time geometry of Minkowski as well as the Euclidean geometry are
continuous geometries. It is true for usual scales of distances. However,
one cannot be sure, that the space-time geometry is continuous in microcosm.
The space-time geometry may appear to be discrete in microcosm. We consider
a discrete space-time geometry and discuss the corollaries of the suggested
discreteness.

The distance function $\rho _{\mathrm{d}}$ of a discrete geometry $\mathcal{G%
}_{\mathrm{d}}$ satisfies the condition%
\begin{equation}
\left\vert \rho _{\mathrm{d}}\left( P,Q\right) \right\vert \notin \left(
0,\lambda _{0}\right) ,\qquad \forall P,Q\in \Omega  \label{a1.1}
\end{equation}%
which means that in the geometry $\mathcal{G}_{\mathrm{d}}$ there are no
distances, which are shorter, than the elementary length $\lambda _{0}$. The
distance $\rho _{\mathrm{d}}\left( P,Q\right) =0$ is admissible. This
condition takes place, if $P=Q$.

Note, that the condition (\ref{a1.1}) is a restriction on the
values of the distance function, but not on values of its argument
(points of $\Omega $), although one considers usually a discrete
geometry as a geometry on a lattice. It is true, that the
geometries on a lattice are discrete geometries, but they form a
very special case of the discrete geometries. Besides, such a
discrete geometry cannot be uniform and isotropic. A general case
of a discrete geometry takes place, when restrictions are imposed
on the admissible values of the world function (distance
function).

The simplest case of a discrete space-time geometry $\mathcal{G}_{\mathrm{d}%
} $ is obtained, if $\mathcal{G}_{\mathrm{d}}=\left\{ \Omega _{\mathrm{M}%
},\sigma _{\mathrm{d}}\right\} $ is given on the manifold $\Omega _{\mathrm{M%
}}$, where the geometry of Minkowski $\mathcal{G}_{\mathrm{M}}=\left\{
\Omega _{\mathrm{M}},\sigma _{\mathrm{M}}\right\} $ is given. The world
function $\sigma _{\mathrm{d}}$ is chosen in the form%
\begin{equation}
\sigma _{\mathrm{d}}\left( P,Q\right) =\sigma _{\mathrm{M}}\left( P,Q\right)
+\frac{1}{2}\lambda _{0}^{2}\mathrm{sgn}\left( \sigma _{\mathrm{M}}\left(
P,Q\right) \right) ,\qquad \forall P,Q\in \Omega _{\mathrm{M}}  \label{a1.2}
\end{equation}%
where $\sigma _{\mathrm{M}}$ is the world function of the geometry of
Minkowski. It is easy to verify, that $\rho _{\mathrm{d}}=\sqrt{2\sigma _{%
\mathrm{d}}}$, defined by (\ref{a1.2}) satisfies the constraint (\ref{a1.1}%
). Such a discrete geometry is uniform and isotropic as well as the geometry
of Minkowski

Besides, in the discrete space-time geometry (\ref{a1.2}) a pointlike
particle cannot be described by a world line, because any world line is a
limit of the broken line, when lengths of its links tend to zero. But in the
discrete geometry $\mathcal{G}_{\mathrm{d}}$ there are no infinitesimal
lengths, and a pointlike particle is described by a world chain (broken
line) instead of continuous world line. Description of a pointlike particle
state by means of the particle position and its momentum becomes inadequate,
because in the continuous space-time geometry the particle 4-momentum $p_{k}$
is described by the relation%
\begin{equation}
p_{k}=g_{kl}\frac{dx^{l}}{d\tau }=g_{kl}\lim_{d\tau \rightarrow 0}\frac{%
x^{l}\left( \tau +d\tau \right) -x^{l}\left( \tau \right) }{d\tau }
\label{a1.3}
\end{equation}%
where $x^{l}=x^{l}\left( \tau \right) $, $l=0,1,2,3$ is an equation of the
world line. The limit in the formula (\ref{a1.3}) does not exist in $%
\mathcal{G}_{\mathrm{d}}$, and the 4-momentum $p_{k}$ is not defined (at any
rate in such a form). In general, the mathematical formalism, based on the
infinitesimal calculus (differential dynamic equations), is inadequate in
the discrete space-time geometry.

The world chain $\mathcal{C}$ of a particle has the form
\begin{equation}
\mathcal{C}:\qquad \dbigcup\limits_{s}\mathbf{P}_{s}\mathbf{P}_{s+1},
\label{a1.4}
\end{equation}%
where vector $\mathbf{P}_{s}\mathbf{P}_{s+1}=\left\{ P_{s},P_{s+1}\right\} $
is an ordered set of two points $P_{s},P_{s+1}$. The lengths $\left\vert
\mathbf{P}_{s}\mathbf{P}_{s+1}\right\vert $ of all vectors are equal
\begin{equation}
\left\vert \mathbf{P}_{s}\mathbf{P}_{s+1}\right\vert =\sqrt{2\sigma _{%
\mathrm{d}}\left( P_{s},P_{s+1}\right) }=\mu ,\qquad s=...0,1,...
\label{a1.5}
\end{equation}%
A new parameter $\mu $ appears in the description of a pointlike
particle. This parameter describes the length of links of the
world chain $\mathcal{C}$. This new parameter (geometrical mass)
may be connected with the particle mass $m$ by means of the
relation
\begin{equation}
m=b\mu  \label{a1.6}
\end{equation}%
where $b$ is some universal constant.

The world chain $\mathcal{C}$ describes the free motion of the pointlike
particle, if in addition to the relation (\ref{a1.5}) the adjacent links of
the chain are parallel%
\begin{equation}
\mathbf{P}_{s}\mathbf{P}_{s+1}\upuparrows \mathbf{P}_{s+1}\mathbf{P}%
_{s+2}:\qquad \left( \mathbf{P}_{s}\mathbf{P}_{s+1}.\mathbf{P}_{s+1}\mathbf{P%
}_{s+2}\right) =\left\vert \mathbf{P}_{s}\mathbf{P}_{s+1}\right\vert \cdot
\left\vert \mathbf{P}_{s+1}\mathbf{P}_{s+2}\right\vert =\mu ^{2}
\label{a1.7}
\end{equation}%
where $\left( \mathbf{P}_{s}\mathbf{P}_{s+1}.\mathbf{P}_{s+1}\mathbf{P}%
_{s+2}\right) $ is the $\sigma $-immanent expression for the scalar product
of two vectors $\mathbf{P}_{s}\mathbf{P}_{s+1}$ and $\mathbf{P}_{s+1}\mathbf{%
P}_{s+2}$.

Relations (\ref{a1.5}) and (\ref{a1.7}) means that adjacent vectors $\mathbf{%
P}_{s}\mathbf{P}_{s+1}$ and $\mathbf{P}_{s+1}\mathbf{P}_{s+2}$ are
equivalent. According to (\ref{a1.5}) the equation (\ref{a1.7}) may written
in the form%
\begin{equation}
\left( \mathbf{P}_{s}\mathbf{P}_{s+1}.\mathbf{P}_{s+1}\mathbf{P}%
_{s+2}\right) =\left\vert \mathbf{P}_{s}\mathbf{P}_{s+1}\right\vert ^{2}=\mu
^{2},\qquad s=....0,1,...  \label{a1.7a}
\end{equation}%
The scalar product of vectors in the discrete geometry $\mathcal{G}_{\mathrm{%
d}}$ has the same form (\ref{c1.2}), but the Euclidean world function $%
\sigma _{\mathrm{E}}$ is replaced by the world function $\sigma _{\mathrm{d}%
} $ of the discrete geometry (\ref{a1.2}). In the limit $\lambda
_{0}\rightarrow 0$ the world function $\sigma _{\mathrm{d}}\rightarrow
\sigma _{\mathrm{M}}$ and the scalar product in $\mathcal{G}_{\mathrm{d}}$
turns to the scalar product in the geometry of Minkowski.

Fixing points $P_{s}$ and $P_{s+1}$ and solving equations (\ref{a1.5}), (\ref%
{a1.7a}) to determine the point $P_{s+2}$, one obtains many solutions for
the point $P_{s+2}$. This circumstance is explained by the fact, that one
has only two equations for determination of four coordinates of the point $%
P_{s+2}$. It means that the links of the world chain (\ref{a1.4}) wobble,
and the shape of the world chain appears to be random (stochastic).

Statistical description of random world chains leads to the Schr\"{o}dinger
equation \cite{R91}, provided the elementary length $\lambda _{0}$ has the
form
\begin{equation}
\lambda _{0}^{2}=\frac{\hbar }{bc}  \label{a1.9}
\end{equation}%
where $\hbar $ is the quantum constant, $c$ is the speed of the light, and $%
b $ is the universal constant, defined by the relation (\ref{a1.6}). The Schr%
\"{o}dinger equation describes a motion of free stochastic (quantum)
particle, and this description contains the particle mass $m$, although the
classical description of a free particle motion does not contain a reference
to the particle mass.

Explanation of quantum effects (the Schr\"{o}dinger equation) as geometrical
effects of the discrete space-time geometry is very important, because it
shows, that introduction of quantum principles is overabundant. This result
suggests to resolve between two paradigms: either discrete space-time
geometry, or continuous space-time geometry equipped by quantum principles.
It is evident, that discrete space-time geometry is the simplest and more
natural solution, than an invention of quantum principles, whose meaning is
unclear. Note, that replacing in quantum mechanics the classical momentum by
an operator or be a matrix, one imitates a multivariance (indeterminacy) of
the particle momentum, described by the relation (\ref{a1.3}) in $\mathcal{G}%
_{\mathrm{d}}$.

Besides, the discrete space-time geometry admits one to describe both the
indeterministic motion of a single particle and the deterministic motion of
an statistically averaged particle, whereas the quantum paradigm admits one
to describe only the statistically averaged particle motion. In application
to the theory of elementary particles the statistically averaged particle
motion does not give a possibility to penetrate in the elementary particle
structure, whereas the single particle motion (even indeterministic) gives a
hope to understand the elementary particles structure.

Formally it means that the conception of particle dynamics in the discrete
space-time geometry distinguishes from the particle dynamics conception in
the space-time geometry of Minkowski, because in the space-time geometry of
Minkowski the formalism of the particle dynamics uses essentially the
continuity of the space-time geometry. This formalism cannot be used in the
discrete space-time geometry.

\section{Description of geometrical objects in physical geometry}

In a physical geometry there is the only basic quantity (world function),
which describes completely all geometric objects of the geometry. All
geometrical objects are described uniformly in all physical geometries. Such
geometrical objects as straight segment and angle are fundamental objects in
conventional presentation of the proper Euclidean geometry. Their properties
are formulated as axioms and they are used at the construction of more
complicate geometric objects. In the physical geometry the straight segment
and the angle are derivative objects, which are constructed in terms of the
world function.

In the $\sigma $-immanent presentation of the proper Euclidean geometry the
straight segment $\mathcal{T}_{\left[ PQ\right] }$ between two points $P$
and $Q$ is described by the relation (\ref{c1.1}). The angle $\varphi $
between two vectors $\mathbf{P}_{0}\mathbf{P}_{1}$ and $\mathbf{Q}_{0}%
\mathbf{Q}_{1}$ is described by the formula%
\begin{equation}
\cos \varphi =\frac{\left( \mathbf{P}_{0}\mathbf{P}_{1}.\mathbf{Q}_{0}%
\mathbf{Q}_{1}\right) }{\left\vert \mathbf{P}_{0}\mathbf{P}_{1}\right\vert
\cdot \left\vert \mathbf{Q}_{0}\mathbf{Q}_{1}\right\vert }=\frac{\sigma
\left( P_{0},Q_{1}\right) +\sigma \left( P_{1},Q_{0}\right) -\sigma \left(
P_{0},Q_{0}\right) -\sigma \left( P_{1},Q_{1}\right) }{\sqrt{4\sigma \left(
P_{0}P_{1}\right) \sigma \left( Q_{0},Q_{1}\right) }}  \label{a2.2}
\end{equation}%
where the scalar product $\left( \mathbf{P}_{0}\mathbf{P}_{1}.\mathbf{Q}_{0}%
\mathbf{Q}_{1}\right) $ is defined by the formula (\ref{c1.2}).

If we try to construct the discrete geometry
$\mathcal{G}_{\mathrm{d}}$, defined by the relation (\ref{a1.1}),
starting from the axioms, which describe properties of the segment
and of the angle, we could not determine, what are properties of
the segment $\mathcal{T}_{\left[ PQ\right] }$ in arbitrary
geometry (for instance, in a discrete geometry). Starting from the
axioms of the proper Euclidean geometry, it is very difficult to
imagine, that the straight segment can be a tube (surface), but
not a one-dimensional line. Starting from these axioms, one cannot
obtain the formula (\ref{c1.1}) for the straight segment in the
discrete space-time geometry. As a result the version of the
discrete space-time geometry was not developed. One developed the
continuous space-time geometry, equipped by quantum principles
instead of the discrete space-time geometry. This approach was
founded on the supposition, that all geometrical objects can be
constructed as combination of fundamental blocks (point, segment,
angle). This supposition means, that a geometry is a logical
construction (axiomatizable geometry). The axiomatizability
condition may be too restrictive for real space-time geometry. At
any rate, it was not reasonable to impose this constraint from the
beginning.

Almost all basic concepts of differential geometry (manifold, dimension,
coordinate system, linear dependence of vectors) are used at the
supposition, that the geometry is continuous. Only concept of distance
function is an exception, which may be defined in a discrete geometry. Using
the distance function as an only basic quantity of a geometry, one may hope
to describe both continuous and discrete geometries. The idea to describe a
geometry in terms of a distance function is an old idea. The following
conditions were imposed on the distance function $\varrho $%
\begin{equation}
\rho \left( P,Q\right) \geq 0,\qquad \forall P,Q\in \Omega  \label{a2.3}
\end{equation}%
\begin{equation}
\rho \left( P,Q\right) =0,\qquad \text{iff }P=Q  \label{a2.4}
\end{equation}%
\begin{equation}
\rho \left( P,Q\right) +\rho \left( Q,R\right) \geq \rho \left( P,R\right)
,\qquad \forall P,Q,R\in \Omega  \label{a2.5}
\end{equation}%
Such a geometry is known as the metric geometry. The condition
(\ref{a2.5}) admits only such geometries, where the straight
segment (\ref{c1.1}) is a one-dimensional point set. The condition
(\ref{a2.3}) forbids an application of the metric geometry to the
space-time.

There were attempts to introduce so called distant geometry \cite{M28,B53},
which is free of restriction (\ref{a2.5}). Note, that the triangle axiom (%
\ref{a2.5}) means, that the straight is an one-dimensional set of points.
However, Blumental \cite{B53} could not overcome the concept of the
Euclidean geometry, that the straight line has no thickness, although a
formal obstruction in the form of the triangle axiom (\ref{a2.5}) has been
overcome. From the logical viewpoint the fact, that in the Euclidean
geometry a straight is an one-dimensional point set, does not mean, that the
straight is a one-dimensional point set in any geometry. Nevertheless,
Blumental introduced a curve as a continuous mapping of interval $(0,1)$
onto the space $\Omega $. He has introduced an additional fundamental
concept (continuous mapping), which cannot be formulated in terms of a
distance. As a result the distant geometry lost its monistic character, when
the geometry is described completely in terms of a distance.

The physical geometry distinguishes from the conventional presentation of
the Euclidean geometry in the relation, that it is a monistic conception,
which is described completely in terms of the world function (or distance).
There is no necessity to reconcile different fundamental concepts of the
geometry, because there is only one fundamental quantity.

A geometrical object is a geometrical image of a physical body. Any
geometrical object is some subset of points in the space-time. However,
geometrical object is not an arbitrary set of points. Geometrical object is
to be defined in the physical geometry in such a way, that similar
geometrical objects (which are images of corresponding physical bodies)
could be recognized in different space-time geometries.

\textit{Definition 2.1:}\emph{\ }A geometrical object $g_{\mathcal{P}%
_{n},\sigma }$ of the geometry $\mathcal{G=}\left\{ \sigma ,\Omega \right\} $
is a subset $g_{\mathcal{P}_{n},\sigma }\subset \Omega $ of the point set $%
\Omega $. This geometrical object $g_{\mathcal{P}_{n},\sigma }$ is a set of
roots $R\in \Omega $ of the function $F_{\mathcal{P}_{n},\sigma }$%
\begin{equation}
F_{\mathcal{P}_{n},\sigma }:\qquad \Omega \rightarrow \mathbb{R}
\label{a2.6}
\end{equation}%
where%
\begin{eqnarray}
F_{\mathcal{P}_{n},\sigma } &:&\quad F_{\mathcal{P}_{n},\sigma }\left(
R\right) =G_{\mathcal{P}_{n},\sigma }\left( u_{1},u_{2},...u_{s}\right)
,\qquad s=\frac{1}{2}\left( n+1\right) \left( n+2\right)  \label{a2.7} \\
u_{l} &=&\sigma \left( w_{i},w_{k}\right) ,\qquad i,k=0,1,...n+1,\qquad
l=1,2,...\frac{1}{2}\left( n+1\right) \left( n+2\right)  \label{a2.8} \\
w_{k} &=&P_{k}\in \Omega ,\qquad k=0,1,...n,\qquad w_{n+1}=R\in \Omega
\label{a2.9}
\end{eqnarray}%
Here $\mathcal{P}_{n}=\left\{ P_{0},P_{1},...,P_{n}\right\} \subset \Omega $
are \ $n+1$ points which are parameters, determining the geometrical object $%
g_{\mathcal{P}_{n},\sigma }$%
\begin{equation}
g_{\mathcal{P}_{n},\sigma }=\left\{ R|F_{\mathcal{P}_{n},\sigma }\left(
R\right) =0\right\} ,\qquad R\in \Omega ,\qquad \mathcal{P}_{n}\in \Omega
^{n+1}  \label{a2.10}
\end{equation}%
$F_{\mathcal{P}_{n},\sigma }\left( R\right) =G_{\mathcal{P}_{n},\sigma
}\left( u_{1},u_{2},...u_{s}\right) $ is an arbitrary function of $\frac{1}{2%
}\left( n+1\right) \left( n+2\right) $ arguments \ $u_{s}$ and of \ $n+1$
parameters $\mathcal{P}_{n}$. The set $\mathcal{P}_{n}$ of the geometric
object parameters will be referred to as the skeleton of the geometrical
object. The subset $g_{\mathcal{P}_{n},\sigma }$ will be referred to as the
envelope of the skeleton. The skeleton is a analog of a frame of reference,
attached rigidly to a physical body. Tracing the skeleton motion, one can
trace the motion of the physical body. When a particle is considered as a
geometrical object, its motion in the space-time is described by the
skeleton $\mathcal{P}_{n}$ motion. At such an approach (the rigid body
approximation) the shape of the envelope is of no importance.

\textit{Remark: }An\textit{\ }arbitrary subset $\Omega ^{\prime }$ of the
point set $\Omega $ is not a geometrical object, in general. It is supposed,
that physical bodies may have only a shape of a geometrical object, because
only in this case one can identify identical physical bodies (geometrical
objects) in different space-time geometries.

Existence of the same geometrical objects in different space-time regions,
having different geometries, arises the question on equivalence of
geometrical objects in different space-time geometries. Such a question was
not arisen before, because one does not consider such a situation, when a
physical body moves from one space-time region to another space-time region,
having another space-time geometry. In general, mathematical technique of
the conventional space-time geometry (differential geometry) is not
applicable for simultaneous consideration of several different geometries of
different space-time regions.

We can perceive the space-time geometry only via motion of physical bodies
in the space-time, or via construction of geometrical objects corresponding
to these physical bodies. As it follows from the \textit{definition 2.1} of
the geometrical object, the function \ $F$ as a function of its arguments
(of world functions of different points) is the same in all physical
geometries. It means, that a geometrical object $\mathcal{O}_{1}$ in the
geometry $\mathcal{G}_{1}=\left\{ \sigma _{1},\Omega _{1}\right\} $ is
obtained from the same geometrical object $\mathcal{O}_{2}$ in the geometry $%
\mathcal{G}_{2}=\left\{ \sigma _{2},\Omega _{2}\right\} $ by means of the
replacement $\sigma _{2}\rightarrow \sigma _{1}$ in the definition of this
geometrical object.

\textit{Definition 2.2: }Geometrical object $g_{P_{n}^{\prime },\sigma
^{\prime }}$ ( $\mathcal{P}_{n}^{\prime }=\left\{ P_{0}^{\prime
},P_{1}^{\prime },..P_{n}^{\prime }\right\} $) in the geometry $\mathcal{G}%
^{\prime }=\left\{ \sigma ^{\prime },\Omega ^{\prime }\right\} $ and the
geometrical object $g_{P_{n},\sigma }$ ( $\mathcal{P}_{n}=\left\{
P_{0},P_{1},..P_{n}\right\} $) in the geometry $\mathcal{G}=\left\{ \sigma
,\Omega \right\} $ are equivalent (equal), if
\begin{equation}
\sigma ^{\prime }\left( P_{i}^{\prime },P_{k}^{\prime }\right) =\sigma
\left( P_{i},P_{k}\right) ,\qquad i,k=0,1,..n  \label{a2.11}
\end{equation}%
and the functions $G_{\mathcal{P}_{n}^{\prime },\sigma ^{\prime }}^{\prime }$
for $g_{P_{n}^{\prime },\sigma ^{\prime }}$ and $G_{\mathcal{P}_{n},\sigma }$
for $g_{P_{n},\sigma }$ in the formula (\ref{a2.7}) coincide
\begin{equation}
G_{\mathcal{P}_{n}^{\prime },\sigma ^{\prime }}^{\prime }\left(
u_{1},u_{2},...u_{s}\right) =G_{\mathcal{P}_{n},\sigma }\left(
u_{1},u_{2},...u_{s}\right)  \label{a2.12}
\end{equation}%
In this case
\begin{equation}
u_{l}\equiv \sigma \left( P_{i},P_{k}\right) =u_{l}^{\prime }\equiv \sigma
^{\prime }\left( P_{i}^{\prime },P_{k}^{\prime }\right) ,\qquad
i,k=0,1,...n,\qquad l=1,2,..n\left( n+1\right) /2  \label{a2.13}
\end{equation}

As far as the physical geometry is determined by its geometrical objects
construction, a physical geometry $\mathcal{G}=\left\{ \sigma ,\Omega
\right\} $ can be obtained from some known standard geometry $\mathcal{G}_{%
\mathrm{st}}=\left\{ \sigma _{\mathrm{st}},\Omega \right\} $ by
means of a deformation of the standard geometry
$\mathcal{G}_{\mathrm{st}}$. Deformation of the standard geometry
$\mathcal{G}_{\mathrm{st}}$ is realized by the replacement $\sigma
_{\mathrm{st}}\rightarrow \sigma $ in all definitions of the
geometrical objects in the standard geometry. The proper Euclidean
geometry is an axiomatizable geometry. It has been constructed by
means of the Euclidean method as a logical construction. The
proper Euclidean geometry is a physical geometry. It may be used
as a standard geometry $\mathcal{G}_{\mathrm{st}}$. Construction
of a physical geometry as a deformation of the proper Euclidean
geometry will be referred to as the deformation principle
\cite{R2007}. The most physical geometries are nonaxiomatizable
geometries. They can be constructed only by means of the
deformation principle.

Description of the elementary particle motion in the space-time contains
only the particle skeleton $\mathcal{P}_{n}=\left\{
P_{0},P_{1},...P_{n}\right\} $. The form of the function (\ref{a2.7}) is of
no importance at the approach, when the particle is considered as a rigid
body. In the elementary particle dynamics only the lengths $\left\vert
\mathbf{P}_{i}\mathbf{P}_{k}\right\vert =\sqrt{2\sigma \left(
P_{i},P_{k}\right) }$ of vectors $\mathbf{P}_{i}\mathbf{P}_{k}$, \ $%
i,k=0,1,...n$ are essential. These vectors are defined by the particle
skeleton $\mathcal{P}_{n}$.

The equivalence $\left( \mathbf{P}_{0}\mathbf{P}_{1}\mathrm{eqv}\mathbf{Q}%
_{0}\mathbf{Q}_{1}\right) $ of two vectors $\mathbf{P}_{0}\mathbf{P}_{1}$
and $\mathbf{Q}_{0}\mathbf{Q}_{1}$ is defined by the relations%
\begin{equation}
\left( \mathbf{P}_{0}\mathbf{P}_{1}\mathrm{eqv}\mathbf{Q}_{0}\mathbf{Q}%
_{1}\right) :\qquad \left( \mathbf{P}_{0}\mathbf{P}_{1}\mathrm{.}\mathbf{Q}%
_{0}\mathbf{Q}_{1}\right) =\left\vert \mathbf{P}_{0}\mathbf{P}%
_{1}\right\vert \cdot \left\vert \mathbf{Q}_{0}\mathbf{Q}_{1}\right\vert
\wedge \left\vert \mathbf{P}_{0}\mathbf{P}_{1}\right\vert =\left\vert
\mathbf{Q}_{0}\mathbf{Q}_{1}\right\vert  \label{b4.5}
\end{equation}%
where%
\begin{equation}
\left\vert \mathbf{P}_{0}\mathbf{P}_{1}\right\vert =\sqrt{2\sigma \left(
P_{0},P_{1}\right) }  \label{b4.6}
\end{equation}%
and the scalar product $\left( \mathbf{P}_{0}\mathbf{P}_{1}\mathrm{.}\mathbf{%
Q}_{0}\mathbf{Q}_{1}\right) $ is defined by the relation (\ref{c1.2})

Skeletons $\mathcal{P}_{n}=\left\{ P_{0},P_{1},...P_{n}\right\} $ and $%
\mathcal{P}_{n}^{\prime }=\left\{ P_{0}^{\prime },P_{1}^{\prime
},...P_{n}^{\prime }\right\} $ may belong to the same geometrical object, if
\begin{equation}
\left\vert \mathbf{P}_{i}\mathbf{P}_{k}\right\vert =\left\vert \mathbf{P}%
_{i}^{\prime }\mathbf{P}_{k}^{\prime }\right\vert ,\qquad i,k=0,1,...n
\label{b4.9}
\end{equation}%
i.e. lengths of all vectors $\mathbf{P}_{i}\mathbf{P}_{k}$ and corresponding
vectors $\mathbf{P}_{i}^{\prime }\mathbf{P}_{k}^{\prime }$ are equal.
However, it is not sufficient for equivalence of skeletons $\mathcal{P}_{n}$
and $\mathcal{P}_{n}^{\prime }$, because the mutual orientation of two
skeletons (corresponding systems of reference) is important for their
equivalence.

Skeletons $\mathcal{P}_{n}=\left\{ P_{0},P_{1},...P_{n}\right\} $ and $%
\mathcal{P}_{n}^{\prime }=\left\{ P_{0}^{\prime },P_{1}^{\prime
},...P_{n}^{\prime }\right\} $ are equivalent, if
\begin{equation}
\left( \mathcal{P}_{n}\mathrm{eqv}\mathcal{P}_{n}^{\prime }\right) :\qquad
\text{if}\ \ \ \ \mathbf{P}_{i}\mathbf{P}_{k}\mathrm{eqv}\mathbf{P}%
_{i}^{\prime }\mathbf{P}_{k}^{\prime },\qquad i,k=0,1,,...n  \label{b.4.10}
\end{equation}%
In other words, the equivalence of skeletons needs equality of the lengths
of vectors $\mathbf{P}_{i}\mathbf{P}_{k}$ and $\mathbf{P}_{i}^{\prime }%
\mathbf{P}_{k}^{\prime }$ and equality of their mutual orientations. For
identification of two geometrical objects one needs only equality of the
lengths of vectors $\mathbf{P}_{i}\mathbf{P}_{k}$ and $\mathbf{P}%
_{i}^{\prime }\mathbf{P}_{k}^{\prime }$

When we start from the world function as an unique fundamental quantity of
the geometry, such geometrical objects as straight segment and the angle
appear as some processing of the fundamental quantity (world function). In
general, one may use another ways of processing of the world function.
However, we use straight segments and skeletons constructed of them, because
we know, how skeletons (systems of reference) are connected with the
particle dynamics. At the free motion of physical body, the skeleton
associated with this body realizes a translational motion. Rotational motion
of a rigid body is not a free motion, because internal forces appear inside
the physical body, and some parts of the body move with an acceleration.

\section{Fluidity of boundary between geometry and \\ dynamics}

Motion of rigid bodies is described by dynamic equations, written in a
space-time geometry. The boundary between the geometry and dynamics is
flexible. One may choose a simple space-time geometry and a complicated
dynamics. On the contrary, one may choose a simple dynamics and a
complicated space-time geometry.

For instance, a motion of a charged particle in a given electromagnetic
field $A_{l}$ is described in the uniform space-time geometry of Minkowski
as a motion under the action of the force field. Corresponding
Hamilton-Jacobi equation has the form%
\[
\left( \frac{\partial S}{\partial x^{i}}-\frac{e}{c}A_{i}\right)
g^{ik}\left( \frac{\partial S}{\partial
x^{k}}-\frac{e}{c}A_{k}\right) =m^{2}c^{2},\quad S=S\left(
x\right) ,\quad g^{ik}=\text{diag}\left\{ c^{-2},-1,-1,-1\right\}
\]%
where $m$ is the particle mass and $e$ is its charge. Such a description
contains two basic essences: space-time geometry and the electromagnetic
field.

The same motion of the charged particle may be described as a free motion of
the particle in the five-dimensional space-time of Kaluza- Klein. In this
case the electromagnetic field is included in the space-time geometry.
Corresponding Hamilton-Jacobi equation has the form%
\[
\frac{\partial S}{\partial X^{A}}\gamma ^{AB}\frac{\partial S}{\partial X^{B}%
}=m_{5}^{2}c^{2},\quad S=S\left( X\right) ,\quad X=\left\{
x^{0},x^{1},x^{2},x^{3},x^{5}\right\} ,\quad \frac{\partial
S}{\partial x^{5}}=\frac{e}{ac}=\text{const}
\]%
\[
\gamma ^{AB}=\left\vert \left\vert
\begin{array}{cc}
g^{ik} & -g^{il}a_{l} \\
-g^{kl}a_{l} & -1+g^{ls}a_{l}a_{s}%
\end{array}%
\right\vert \right\vert ,\qquad a_{l}=aA_{l},\qquad m_{5}^{2}=m^{2}-\left(
\frac{e}{ac}\right) ^{2}
\]%
where $a$ is some constant.

Such a  description of the space-time geometry "absorbs" dynamics. Dynamics
becomes simpler, but geometry becomes more complicated. Geometry ceases to
be uniform, because it contains information on the electromagnetic field.

If the space-time geometry is described by the only fundamental quantity
(world function), and the particle motion is supposed to be free, such
"absorption" of dynamics by the space-time geometry leads to a
simplification of the particle dynamics. As a result a monistic description
of the particle motion in terms of the world function arises.

In the microcosm, where several different force fields may exist, such a
reduction of dynamics to the space-time geometry may strongly simplify the
particle motion description. If there are several force fields, their
reconciliation is very complicated, because the number of reconciling
variants increases rapidly with the number of force fields. In the case,
when all fields are absorbed by the space-time geometry, one has a monistic
conception, which does not need any reconciling. Of course, the world
function becomes more complicated, however it is only one invariant
quantity, which is a function of two space-time points. It is an attractive
circumstance, that the particle dynamics may be formulated in coordinateless
form and in invariant terms.

As we shall see, a skeleton conception of particle dynamics is formulated
very simply in the form of difference (not differential) equations. It is
quite natural, because the space-time geometry may appear to be discrete. In
the discrete space-time geometry, where there are no infinitesimal
distances, the differential equation may not be applicable.

However, in the case, when the characteristic scale is much more, than the
elementary length, one may use differential equations instead of difference
ones.

\section{Discreteness and its manifestations}

The simplest discrete space-time geometry $\mathcal{G}_{\mathrm{d}}$ is
described by the world function (\ref{a1.2}). Density of points in $\mathcal{%
G}_{\mathrm{d}}$ with respect to point density in $\mathcal{G}_{\mathrm{M}}$
is described by the relation%
\begin{equation}
\frac{d\sigma _{\mathrm{M}}}{d\sigma _{\mathrm{d}}}=\left\{
\begin{array}{ccc}
0 & \text{if} & \left\vert \sigma _{\mathrm{d}}\right\vert <\frac{1}{2}%
\lambda _{0}^{2} \\
1 & \text{if} & \left\vert \sigma _{\mathrm{d}}\right\vert >\frac{1}{2}%
\lambda _{0}^{2}%
\end{array}%
\right.  \label{a5.1}
\end{equation}

If the world function has the form
\begin{equation}
\sigma _{\mathrm{g}}=\sigma _{\mathrm{M}}+\frac{\lambda _{0}^{2}}{2}\left\{
\begin{array}{ccc}
\mathrm{sgn}\left( \sigma _{\mathrm{M}}\right) & \text{if} & \left\vert
\sigma _{\mathrm{M}}\right\vert >\sigma _{0} \\
\frac{\sigma _{\mathrm{M}}}{\sigma _{0}} & \text{if} & \left\vert \sigma _{%
\mathrm{M}}\right\vert \leq \sigma _{0}%
\end{array}%
\right.  \label{a5.2}
\end{equation}%
where $\sigma _{0}=$const, $\sigma _{0}\geq 0$, the relative density of
points has the form%
\begin{equation}
\frac{d\sigma _{\mathrm{M}}}{d\sigma _{\mathrm{g}}}=\left\{
\begin{array}{ccc}
\frac{2\sigma _{0}}{2\sigma _{0}+\lambda _{0}^{2}} & \text{if} & \left\vert
\sigma _{\mathrm{g}}\right\vert <\sigma _{0}+\frac{1}{2}\lambda _{0}^{2} \\
1 & \text{if} & \left\vert \sigma _{\mathrm{g}}\right\vert >\sigma _{0}+%
\frac{1}{2}\lambda _{0}^{2}%
\end{array}%
\right.  \label{a5.3}
\end{equation}%
If the parameter $\sigma _{0}\rightarrow 0$, the world function $\sigma _{%
\mathrm{g}}\rightarrow \sigma _{\mathrm{d}}$ and the point density (\ref%
{a5.3}) tends to the point density (\ref{a5.1}). The space-time geometry $%
\mathcal{G}_{\mathrm{g}}$, described by the world function $\sigma _{\mathrm{%
g}}$ is a geometry, which is a partly discrete geometry, because it is
intermediate between the discrete geometry $\mathcal{G}_{\mathrm{d}}$ and
the continuous geometry $\mathcal{G}_{\mathrm{M}}$. We shall refer to the
geometry $\mathcal{G}_{\mathrm{g}}$ as a granular geometry.

Deflection of the discrete space-time geometry from the continuous geometry
of Minkowski generates special properties of the geometry, which are
corollaries of impossibility of the linear vector space introduction.

Let $\mathbf{Q}_{0}\mathbf{Q}_{1}$ be a timelike vector in $\mathcal{G}_{%
\mathrm{d}}$ ($\sigma _{\mathrm{d}}(Q_{0},Q_{1})>0$). We try to
determine a vector $\mathbf{P}_{0}\mathbf{P}_{1}$ at the point
$P_{0}$, which is equivalent to vector
$\mathbf{Q}_{0}\mathbf{Q}_{1}$. Let for simplicity coordinates has
the form
\begin{equation}
P_{0}=Q_{0}=\left\{ 0,0,0,0\right\} ,\qquad Q_{1}=\left\{ 1,0,0,0\right\}
,\qquad P_{1}=\left\{ x^{0},\mathbf{x}\right\} =\left\{
x^{0},x^{1},x^{2},x^{3}\right\}  \label{a5.4}
\end{equation}%
In this coordinate system the world function of geometry Minkowski has the
form%
\begin{equation}
\sigma _{\mathrm{M}}\left( x,x^{\prime }\right) =\frac{1}{2}\left( \left(
x^{0}-x^{0\prime }\right) ^{2}-\left( \mathbf{x}-\mathbf{x}^{\prime }\right)
^{2}\right)  \label{a5.5}
\end{equation}%
and $\sigma _{\mathrm{d}}$ is determined by the relation (\ref{a1.2}). We
are to determine coordinates $x$ of the point $P_{1}$ from two equations (%
\ref{b4.5}), which have the form%
\begin{equation}
\frac{1}{2}\left( \left( \left( x^{0}\right) ^{2}-\mathbf{x}^{2}\right) +%
\frac{\lambda _{0}^{2}}{2}\right) =\frac{1}{2}\left( 1+\frac{\lambda _{0}^{2}%
}{2}\right)  \label{a5.6}
\end{equation}%
\begin{equation}
\frac{1}{2}\left( 1+\frac{\lambda _{0}^{2}}{2}\right) +\frac{1}{2}\left(
\left( \left( x^{0}\right) ^{2}-\mathbf{x}^{2}\right) +\frac{\lambda _{0}^{2}%
}{2}\right) -\frac{1}{2}\left( \left( \left( x^{0}-1\right) ^{2}-\mathbf{x}%
^{2}\right) +\frac{\lambda _{0}^{2}}{2}\right) =\left( 1+\frac{\lambda
_{0}^{2}}{2}\right)  \label{a5.7}
\end{equation}%
After simplification one obtains%
\begin{equation}
x^{0}=1+\frac{\lambda _{0}^{2}}{4},\qquad \mathbf{x}^{2}=\left( 1+\frac{%
\lambda _{0}^{2}}{4}\right) ^{2}-1  \label{a5.8}
\end{equation}%
If $\lambda _{0}=0$, then the discrete geometry turns to the geometry of
Minkowski, and $\mathbf{x}^{2}=0$. Three equations
\begin{equation}
x^{1}=0,\quad x^{2}=0,\quad x^{3}=0  \label{5.10}
\end{equation}
follow from one equation $\mathbf{x}^{2}=0$. It means, that the geometry of
Minkowski is a degenerate geometry, because different solutions of the
discrete geometry merge into one solution of the geometry of Minkowski.

One obtains for coordinates of point $P_{1}$:%
\begin{equation}
P_{1}=\left\{ \sqrt{r+1},r\sin \theta ,r\sin \theta \sin \varphi ,r\sin
\theta \cos \varphi \right\} ,\qquad r=\frac{\lambda _{0}}{\sqrt{2}}\sqrt{1+%
\frac{\lambda _{0}^{2}}{4}}  \label{a5.9}
\end{equation}%
where the angles $\theta $, $\varphi $ are arbitrary. Such a situation is
formulated as multivariance of the equivalency relation. All solutions lie
on the surface of two-dimensional sphere of radius $r\simeq \lambda _{0}/%
\sqrt{2}$ ($\lambda _{0}^{2}\ll 1$).

Let now $\mathbf{Q}_{0}\mathbf{Q}_{1}$ be a spacelike vector in $\mathcal{G}%
_{\mathrm{d}}$ ($\sigma _{\mathrm{d}}(Q_{0},Q_{1})<0$). We try to determine
a vector $\mathbf{P}_{0}\mathbf{P}_{1}$ at the point $P_{0}$, which is
equivalent to vector $\mathbf{Q}_{0}\mathbf{Q}_{1}$. Let
\begin{equation}
P_{0}=Q_{0}=\left\{ 0,0,0,0\right\} ,\qquad Q_{1}=\left\{ 0,1,0,0\right\}
,\qquad P_{1}=\left\{ x^{0},\mathbf{x}\right\} =\left\{
x^{0},x^{1},x^{2},x^{3}\right\}  \label{5.11}
\end{equation}%
Two equations (\ref{b4.5}) have the form%
\begin{equation}
\frac{1}{2}\left( \left( x^{0}\right) ^{2}-\mathbf{x}^{2}\right) -\frac{%
\lambda _{0}^{2}}{2}=-\frac{1}{2}-\frac{\lambda _{0}^{2}}{2}  \label{a5.12}
\end{equation}%
\begin{eqnarray}
&&\left( -\frac{1}{2}-\frac{\lambda _{0}^{2}}{2}\right)
+\frac{1}{2}\left(
\left( x^{0}\right) ^{2}-\mathbf{x}^{2}\right) -\frac{\lambda _{0}^{2}}{2}-%
\frac{1}{2}\left( \left( x^{0}\right) ^{2}-\left( x^{1}-1\right)
^{2}-\left( x^{2}\right) ^{2}-\left( x^{3}\right) ^{2}\right)
\nonumber \\
&=& -\frac{\lambda _{0}^{2}}{2} -\left( 1+\lambda _{0}^{2}\right)
\label{a5.13}
\end{eqnarray}%
After simplification one obtains%
\begin{equation}
x^{1}=1+\frac{\lambda _{0}^{2}}{2},\qquad \left( x^{0}\right) ^{2}-\mathbf{x}%
^{2}=-1  \label{a5.15}
\end{equation}%
The solution for the point $P_{1}$ has the form%
\begin{equation}
P_{1}=\left\{ \sqrt{a_{2}^{2}+a_{3}^{2}+\frac{\lambda _{0}^{2}}{2}\left( 2+%
\frac{\lambda _{0}^{2}}{2}\right) },1+\frac{\lambda _{0}^{2}}{2}%
,a_{2},a_{3}\right\}  \label{a5.16}
\end{equation}%
where $a_{2}$ and $a_{3}$ are arbitrary real numbers.\ In the case of
spacelike vectors $\mathbf{Q}_{0}\mathbf{Q}_{1}$, $\mathbf{P}_{0}\mathbf{P}%
_{1}$ the solution is not unique, even if $\lambda _{0}=0$, and geometry is
the geometry of Minkowski. All solutions lie on the surface of
two-dimensional sphere of arbitrary radius $\sqrt{a_{2}^{2}+a_{3}^{2}}$.

Thus, both the discrete geometry and the geometry of Minkowski are
multivariant with respect to spacelike vectors. However, this circumstance
remains to be unnoticed in the conventional relativistic particle dynamics,
because the spacelike vectors do not used there.

Multivariance of the discrete geometry leads to intransitivity of the
equivalence relation of two vectors. Indeed, if $\left( \mathbf{Q}_{0}%
\mathbf{Q}_{1}\mathrm{eqv}\mathbf{P}_{0}\mathbf{P}_{1}\right) $ and $\left(
\mathbf{Q}_{0}\mathbf{Q}_{1}\mathrm{eqv}\mathbf{P}_{0}\mathbf{P}_{1}^{\prime
}\right) $, but vector $\left( \mathbf{P}_{0}\mathbf{P}_{1}\overline{\mathrm{%
eqv}}\mathbf{P}_{0}\mathbf{P}_{1}^{\prime }\right) $, it means
intransitivity of the equivalence relation. Besides, it means that the
discrete geometry is nonaxiomatizable, because in any logical construction
the equivalence relation is transitive.

Transport of a vector $\mathbf{P}_{0}\mathbf{P}_{1}$ to some point $Q_{0}$
leads to some indeterminacy of the result of this transport, because at the
point $Q_{0}$ there are many vectors $\mathbf{Q}_{0}\mathbf{Q}_{1}$, $%
\mathbf{Q}_{0}\mathbf{Q}_{1}^{\prime }$,..., which are equivalent to the
vector $\mathbf{P}_{0}\mathbf{P}_{1}.$

Sum $\mathbf{Q}_{0}\mathbf{S}$ of two vectors $\mathbf{Q}_{0}\mathbf{Q}_{1}$%
\textrm{\ }and $\mathbf{Q}_{1}\mathbf{S}$, when the end of one vector is an
origin of the other, is defined by points $Q_{0}$ and $S$%
\begin{equation}
\mathbf{Q}_{0}\mathbf{S=Q}_{0}\mathbf{Q}_{1}+\mathbf{Q}_{1}\mathbf{S}
\label{a5.14}
\end{equation}

Sum $\mathbf{Q}_{0}\mathbf{S}$ of two vectors $\mathbf{Q}_{0}\mathbf{Q}_{1}$%
\textrm{\ }and $\mathbf{P}_{0}\mathbf{P}_{1}$ at the point $Q_{0}$ is
defined by the relations%
\begin{equation}
\mathbf{Q}_{0}\mathbf{S=Q}_{0}\mathbf{Q}_{1}+\mathbf{Q}_{1}\mathbf{S,\qquad }%
\left( \mathbf{Q}_{1}\mathbf{S}\mathrm{eqv}\mathbf{P}_{0}\mathbf{P}%
_{1}\right)  \label{a5.17}
\end{equation}%
In the discrete geometry the sum of two vectors is not unique, in general.

Result of multiplication of a vector $\mathbf{P}_{0}\mathbf{P}_{1}$ by a
real number $a$ is not unique also. The result $\mathbf{P}_{0}\mathbf{S}$ of
such a multiplication by a number $a$ is defined by relations
\begin{equation}
\mathbf{P}_{0}\mathbf{P}_{1}\upuparrows \mathbf{P}_{0}\mathbf{S\wedge }%
\left\vert \mathbf{P}_{0}\mathbf{S}\right\vert =a\left\vert \mathbf{P}_{0}%
\mathbf{P}_{1}\right\vert  \label{a5.18}
\end{equation}%
or in terms of algebraic relations%
\begin{equation}
\left( \left( \mathbf{P}_{0}\mathbf{P}_{1}.\mathbf{P}_{0}\mathbf{S}\right)
=a\left\vert \mathbf{P}_{0}\mathbf{P}_{1}\right\vert \cdot \left\vert
\mathbf{P}_{0}\mathbf{P}_{1}\right\vert \right) \mathbf{\wedge }\left\vert
\mathbf{P}_{0}\mathbf{S}\right\vert =a\left\vert \mathbf{P}_{0}\mathbf{P}%
_{1}\right\vert  \label{a5.19}
\end{equation}

Thus, results of vectors summation and of a multiplication of a vector by a
real number are not unique, in general, in the discrete geometry. It means,
that one cannot introduce a linear vector space in the discrete geometry.

Let the discrete geometry is described by $n$ coordinates. Let the skeleton $%
\mathcal{P}_{n}=\left\{ P_{0},P_{1},...P_{n}\right\} $ determine $n$ vectors
$\mathbf{P}_{0}\mathbf{P}_{k}$,\textbf{\ }$k=1,2,...n$, which are linear
independent in the sense
\begin{equation}
F_{n}\left( \mathcal{P}_{n}\right) =\det \left\Vert \left( \mathbf{P}_{0}%
\mathbf{P}_{i}.\mathbf{P}_{0}\mathbf{P}_{k}\right) \right\Vert \neq 0\qquad
i,k=1,2,...n  \label{a5.20}
\end{equation}%
One can determine uniquely projections of a vector $\mathbf{Q}_{0}\mathbf{Q}%
_{1}$ onto vectors $\mathbf{P}_{0}\mathbf{P}_{k}$,\textbf{\ }$k=1,2,...n$ by
means of relations
\begin{equation}
\Pr \left( \mathbf{Q}_{0}\mathbf{Q}_{1}\right) _{\mathbf{P}_{0}\mathbf{P}%
_{k}}=\frac{\left( \mathbf{Q}_{0}\mathbf{Q}_{1}.\mathbf{P}_{0}\mathbf{P}%
_{k}\right) }{\left\vert \mathbf{P}_{0}\mathbf{P}_{k}\right\vert }
\label{a5.21}
\end{equation}%
However, one cannot reestablish the vector $\mathbf{Q}_{0}\mathbf{Q}_{1}$,
using its projections onto vectors $\mathbf{P}_{0}\mathbf{P}_{k}$. Thus, all
operations of the linear vector space are not unique in the discrete
geometry.

Mathematical technique of continuous geometry is not adequate for
application in a discrete geometry, because it is too special and adapted
for continuous geometry.

\section{Skeleton conception of particle dynamics}

Let us suppose, that the particle motion is free, and all force fields are
included in the space-time geometry. The particle motion is described by the
motion of the particle skeleton $\mathcal{P}_{n}=\left\{
P_{0},P_{1},...P_{n}\right\} $, which is a set of $n+1$ space-time points.
The skeleton is an analog of the frame of reference attached rigidly to the
particle (physical body). Tracing the skeleton motion, one traces the
particle motion. The skeleton motion is described by a world chain $\mathcal{%
C}$ of connected skeletons
\begin{equation}
\mathcal{C=}\dbigcup\limits_{s=-\infty }^{s=+\infty }\mathcal{P}_{n}^{\left(
s\right) }  \label{a4.1}
\end{equation}%
Skeletons $\mathcal{P}_{n}^{\left( s\right) }$ of the world chain are
connected in the sense, that
\begin{equation}
P_{1}^{\left( s\right) }=P_{0}^{\left( s+1\right) },\qquad s=...0,1,...
\label{a4.2}
\end{equation}%
The vector $\mathbf{P}_{0}^{\left( s\right) }\mathbf{P}_{1}^{\left( s\right)
}=\mathbf{P}_{0}^{\left( s\right) }\mathbf{P}_{0}^{\left( s+1\right) }$ is
the leading vector, which determines the direction of the world chain. If
the particle motion is free, the adjacent skeletons are equivalent%
\begin{equation}
\mathcal{P}_{n}^{\left( s\right) }\text{eqv}\mathcal{P}_{n}^{\left(
s+1\right) }:\qquad \mathbf{P}_{i}^{\left( s\right) }\mathbf{P}_{k}^{\left(
s\right) }\text{eqv}\mathbf{P}_{i}^{\left( s+1\right) }\mathbf{P}%
_{k}^{\left( s+1\right) },\qquad i,k=0,1,...n,\qquad s=..0,1,..  \label{a4.3}
\end{equation}%
If the particle is described by the skeleton $\mathcal{P}_{n}^{\left(
s\right) }$, the world chain (\ref{a4.1}) has $n(n+1)/2$ invariants%
\begin{equation}
\mu _{ik}=\left\vert \mathbf{P}_{i}^{\left( s\right) }\mathbf{P}_{k}^{\left(
s\right) }\right\vert ^{2}=2\sigma \left( P_{i}^{\left( s\right)
},P_{k}^{s}\right) ,\qquad i,k=0,1,...n,\qquad s=...0,1,...  \label{a4.4}
\end{equation}%
which are constant along the whole world chain.

Equations (\ref{a4.3}) form a system of $n\left( n+1\right) $ difference
equations for determination of $nD$ coordinates of $n$ skeleton points $%
\left\{ P_{1},P_{2},..P_{n}\right\} $, where $D$ is the dimension of the
space-time.

In the case of pointlike particle, when $n=1,$ $D=4$, the number of
equations $n_{e}=2$, whereas the number of variables $n_{v}=4$. The number
of equations is less, than the number of dynamic variables. In the discrete
space-time geometry (\ref{a1.2}) position of the adjacent skeleton is not
uniquely determined. As a result the world chain wobbles. In the
nonrelativistic approximation a statistical description of the stochastic
world chains leads to the Schr\"{o}dinger equations \cite{R91}, if the
elementary length $\lambda _{0}\ $has the form (\ref{a1.9}).

Dynamic equations (\ref{a4.4}) are difference equations. At the large scale,
when one may go to the limit $\lambda _{0}=0$, the dynamic equations (\ref%
{a4.4}) turn to the differential dynamic equations. In the case of pointlike
particle $(n=1)$ and of the Kaluza-Klein five-dimensional space-time
geometry these equation describe the motion of a charged particle in the
given electromagnetic field.

Dynamic equations (\ref{a4.3}) realize the skeleton conception of
particle dynamics in the microcosm. The skeleton conception of
dynamics distinguishes from the conventional conception of
particle dynamics in the relation that the number of dynamic
equations may differ from the number of dynamic variables, which
are to be determined. In the conventional conception of particle
dynamics the number of dynamic equations (first order) coincide
always with the number of dynamic variables, which are to be
determined. As a result the motion of a particle (or of a averaged
particle) appears to be deterministic. In the case of quantum
particles, whose motion is stochastic (indeterministic), the
dynamical equations are written for a statistical ensemble of
indeterministic particles (or for the statistically averaged
particle).

Statistical ensemble of indeterministic particles is a continuous dynamic
system. It may be considered as a fluid, whose state is described by
functions of spatial coordinates. These functions may be hydrodynamic
variables: density $\rho \left( t,\mathbf{x}\right) $ and velocity $\mathbf{v%
}\left( t,\mathbf{x}\right) $, or wave function $\psi \left( t,\mathbf{x}%
\right) $, which is also a method of the fluid description \cite{R1999}. In
any case the statistically averaged particle (or statistical ensemble) has
infinite number of the freedom degrees. Nevertheless the dynamic equations
are deterministic. They can be derived from a variational principle.

Dynamic equations (\ref{a4.3}) cannot be derived from a variational
principle, because evolution of the particle skeleton may be
indeterministic. Let us illustrate the difference between the skeleton
conception of particle dynamics and the conventional conception of particle
dynamics in the example of stochastic particles $\mathcal{S}_{\mathrm{st}}$
\cite{R2010}

The action for the statistical ensemble $\mathcal{E}\left[ \mathcal{S}_{%
\mathrm{st}}\right] $ is written in the form%
\begin{equation}
\mathcal{A}_{\mathcal{E}\left[ \mathcal{S}_{\mathrm{st}}\right] }\left[
\mathbf{x},\mathbf{u}\right] =\int \dint\limits_{V_{\xi }}\left\{ \frac{m}{2}%
\mathbf{\dot{x}}^{2}+\frac{m}{2}\mathbf{u}^{2}-\frac{\hbar }{2}\mathbf{%
\nabla u}\right\} \rho _{0}\left( \mathbf{\xi }\right) dtd\mathbf{\xi }%
,\qquad \mathbf{\dot{x}\equiv }\frac{d\mathbf{x}}{dt}  \label{d1.5}
\end{equation}%
The variable $\mathbf{x}=\mathbf{x}\left( t,\mathbf{\xi }\right) $ describes
the regular component of the particle motion. The variable $\mathbf{u}=%
\mathbf{u}\left( t,\mathbf{x}\right) $ describes the mean value of the
stochastic velocity component, $\hbar $ is the quantum constant. The second
term in (\ref{d1.5}) describes the kinetic energy of the stochastic velocity
component. The third term describes interaction between the stochastic
component $\mathbf{u}\left( t,\mathbf{x}\right) $ and the regular component $%
\mathbf{\dot{x}}\left( t,\mathbf{\xi }\right) $. The variable $\mathbf{\xi =}%
\left\{ \xi _{1},\xi _{2},\xi _{3}\right\} $ labels the elements of the
statistical ensemble. The operator
\begin{equation}
\mathbf{\nabla =}\left\{ \frac{\partial }{\partial x^{1}},\frac{\partial }{%
\partial x^{2}},\frac{\partial }{\partial x^{3}}\right\}  \label{d1.5a}
\end{equation}%
is defined in the space of coordinates $\mathbf{x}$. Dynamic equations for
the dynamic system $\mathcal{E}\left[ \mathcal{S}_{\mathrm{st}}\right] $ are
obtained as a result of variation of the action (\ref{d1.5}) with respect to
dynamic variables $\mathbf{x}$ and $\mathbf{u}$.

To obtain the action functional for $\mathcal{S}_{\mathrm{st}}$ from the
action (\ref{d1.5}) for $\mathcal{E}\left[ \mathcal{S}_{\mathrm{st}}\right] $%
, we should omit integration over $\mathbf{\xi }$ in (\ref{d1.5}). We obtain%
\begin{equation}
\mathcal{A}_{\mathcal{S}_{\mathrm{st}}}\left[ \mathbf{x},\mathbf{u}\right]
=\int \left\{ \frac{m}{2}\mathbf{\dot{x}}^{2}+\frac{m}{2}\mathbf{u}^{2}-%
\frac{\hbar }{2}\mathbf{\nabla u}\right\} dt,\qquad \mathbf{\dot{x}\equiv }%
\frac{d\mathbf{x}}{dt}  \label{d1.6}
\end{equation}%
where $\mathbf{x}=\mathbf{x}\left( t\right) $ and $\mathbf{u}=\mathbf{u}%
\left( t,\mathbf{x}\right) $ are dependent dynamic variables. The action
functional (\ref{d1.6}) is not well defined (for $\hbar \neq 0$), because
the operator $\mathbf{\nabla }$ is defined in some 3-dimensional vicinity of
point $\mathbf{x}$, but not at the point $\mathbf{x}$ itself. As far as the
action functional (\ref{d1.6}) is not well defined, one cannot obtain
dynamic equations for $\mathcal{S}_{\mathrm{st}}$. By definition it means
that the particle $\mathcal{S}_{\mathrm{st}}$ is stochastic. Setting $\hbar
=0$ in (\ref{d1.6}), we transform the action (\ref{d1.6}) into the action
for deterministic particle $\mathcal{S}_{\mathrm{d}}$%
\begin{equation}
\mathcal{A}_{\mathcal{S}_{\mathrm{d}}}\left[ \mathbf{x}\right] =\int \frac{m%
}{2}\mathbf{\dot{x}}^{2}dt,\qquad \mathbf{\dot{x}\equiv }\frac{d\mathbf{x}}{%
dt}  \label{d1.1}
\end{equation}%
because in this case $\mathbf{u}=0$ in virtue of dynamic equations.

After proper change of dynamic variables the action (\ref{d1.5}) is
transformed to the form (see details of transformation in \cite{R2010})
\begin{equation}
\mathcal{A}[\psi ,\psi ^{\ast }]=\int \left\{ \frac{i\hbar }{2}(\psi ^{\ast
}\partial _{0}\psi -\partial _{0}\psi ^{\ast }\cdot \psi )-\frac{\hbar ^{2}}{%
2m}\mathbf{\nabla }\psi ^{\ast }\cdot \mathbf{\nabla }\psi \right. +\left.
\frac{\hbar ^{2}}{8m}\rho \mathbf{\nabla }s_{\alpha }\mathbf{\nabla }%
s_{\alpha }\right\} \mathrm{d}^{4}x  \label{d6.10}
\end{equation}%
where $\psi $ is a two-component complex wave function%
\begin{equation}
\rho =\psi ^{\ast }\psi ,\qquad s_{\alpha }=\frac{\psi ^{\ast }\sigma
_{\alpha }\psi }{\rho },\qquad \alpha =1,2,3  \label{d6.7}
\end{equation}%
$\sigma _{\alpha }$ are $2\times 2$ Pauli matrices%
\begin{equation}
\sigma _{1}=\left(
\begin{array}{cc}
0 & 1 \\
1 & 0%
\end{array}%
\right) ,\qquad \sigma _{2}=\left(
\begin{array}{cc}
0 & -i \\
i & 0%
\end{array}%
\right) ,\qquad \sigma _{3}=\left(
\begin{array}{cc}
1 & 0 \\
0 & -1%
\end{array}%
\right) ,  \label{d6.8}
\end{equation}

In the case, when the wave function $\psi $ is one-component, for instance $%
\psi =\left\{ _{0}^{\psi _{1}}\right\} $, the quantities $\mathbf{s=}\left\{
s_{1},s_{2},s_{3}\right\} $ are constant ($s_{1}=0,\ \ s_{2}=0,\ \ s_{3}=1$%
), the action (\ref{d6.10}) turns into
\begin{equation}
\mathcal{A}[\psi ,\psi ^{\ast }]=\int \left\{ \frac{i\hbar }{2}(\psi ^{\ast
}\partial _{0}\psi -\partial _{0}\psi ^{\ast }\cdot \psi )-\frac{\hbar ^{2}}{%
2m}\mathbf{\nabla }\psi ^{\ast }\cdot \mathbf{\nabla }\psi \right\} \mathrm{d%
}^{4}x  \label{d6.11}
\end{equation}%
The dynamic equation, generated by the action (\ref{d6.11}), is the Schr\"{o}%
dinger equation%
\begin{equation}
i\hbar \partial _{0}\psi +\frac{\hbar ^{2}}{2m}\mathbf{\nabla }^{2}\psi =0
\label{d6.12}
\end{equation}%
This dynamic equation describes the irrotational flow of the fluid.

In the general case one obtains from the action (\ref{d6.10})%
\begin{equation}
()i\hbar \partial _{0}\psi +\frac{\hbar ^{2}}{2m}\mathbf{\nabla }^{2}\psi +%
\frac{\hbar ^{2}}{8m}\mathbf{\nabla }^{2}s_{\alpha }\cdot \left( s_{\alpha
}-2\sigma _{\alpha }\right) \psi -\frac{\hbar ^{2}}{4m}\frac{\mathbf{\nabla }%
\rho }{\rho }\mathbf{\nabla }s_{\alpha }\sigma _{\alpha }\psi =0
\label{d6.15}
\end{equation}%
where two last terms describe the vorticity of the fluid flow.

In the conventional conception of dynamics one can obtain dynamic equation
for the statistically averaged particle (i.e. statistical ensemble
normalized to one particle), but there are no dynamic equations for a single
stochastic particle. In the skeleton conception of dynamics there are
dynamic equations for a single particle. These equations are many-valued
(multivariant), but they do exist. One can derive dynamic equations for the
statistically averaged particle which is a kind of a fluid (continuous
medium). A flow of this fluid is deterministic, and dynamic equations for
this flow can be obtained from a variational principle.

Thus, a difference between the conventional conception of dynamics
and the skeleton conception of dynamics lies in the description of
a single particle. Conventional conception cannot obtain dynamic
equations for a single stochastic particle, whereas skeleton
conception can obtain the dynamic equations for a single particle,
although these equations are multivariant. This difference is
conditioned by the fact, that in the case of the skeleton
conception the stochasticity is conditioned by the space-time
geometry, whereas in the conventional conception of dynamics the
stochasticity is introduced axiomatically, and there is no model
of the stochasticity.

One may compare this situation with the situation in description
of thermal phenomena. If we use the axiomatic thermodynamics, we
cannot say nothing on the structure of gas molecules. We cannot
say anything even on the existence of molecules. If we use the
statistical physics we obtain some information on behavior and
structure of gas molecules, although their motion is chaotic and
multivariant.

The skeleton conception of the particle dynamics realizes a more detailed
description of elementary particle, one may hope to obtain some information
on the elementary particle structure.

We have now only two examples of the skeleton conception application.
Considering compactification in the 5-dimensional discrete space-time
geometry of Kaluza-Klein, and imposing condition of uniqueness of the world
function, one obtains that the value of the electric charge of a stable
elementary particle is restricted by the elementary charge \cite{R2011}.
This result has been known experimentally, but it could not be explained
theoretically, because in the continuous space-time geometry nobody
considers the world function as a fundamental quantity and demand its
uniqueness.

Another example concerns structure of Dirac particles (fermions).
Consideration in framework of skeleton conception \cite{R2008} shows, that
world chain of a fermion is a spacelike helix with timelike axis. The
averaged world chain of a free fermion is a timelike straight line. The
helical motion of a skeleton generates an angular moment (spin) and magnetic
moment. Such a result looks rather reasonable. In the conventional
conception of the particle dynamics the spin and magnetic moment of a
fermion are postulated without a reference to its structure.

\section{Concluding remarks}

Thus, the supposition on the space-time geometry discreteness seems to be
more natural and reasonable, than the supposition on quantum nature of
physical phenomena in microcosm. Discreteness is simply a property of the
space-time, whereas quantum principles assume introduction of new essences.

Formalism of the discrete geometry is very simple. It does not contain
theorems with complicated proofs. Nevertheless the discrete geometry and its
formalism is perceived hardly. The discrete geometry was not developed in
the twentieth century, although the discrete space-time was necessary for
description of physical phenomena in microcosm. It was rather probably, that
the space-time is discrete in microcosm. What is a reason of the discrete
geometry disregard? We try to answer this important question.

The discrete geometry was not developed, because it could be obtained only
as a generalization of the proper Euclidean geometry. But almost all
concepts and quantities of the proper Euclidean geometry use essentially
concepts of the continuous geometry. They could not be used for construction
of a discrete geometry. Only world function (or distance) does not use a
reference to the geometry continuity. Only coordinateless expressions (\ref%
{c1.1}) - (\ref{c1.5}) of basic quantities of the Euclidean geometry in
terms of world function admit one to construct a discrete geometry and other
physical geometries.

Supposition, that any geometry is to be axiomatizable, was the second
obstacle on the way of the discrete geometry construction. In general, the
fact, that the proper Euclidean geometry is a degenerate geometry, was
another obstacle. In particular, being a physical geometry, the proper
Euclidean geometry is an axiomatizable geometry, and this circumstance is an
evidence of its degeneracy. It is very difficult to obtain a general
conception as a generalization of a degenerate conception, because many
quantities of the general conception coincide in the degenerate conception.
It is rather difficult to disjoint them. For instance, a physical geometry
is multivariant, in general. Single-variant physical geometry is a
degenerate geometry. In the physical geometry the straight segment (\ref%
{c1.1}) is a surface (tube), in general. In the degenerate physical geometry
(the proper Euclidean geometry) the straight segment is a one-dimensional
line. How can one guess, that a straight segment is a surface, in general?
Besides, multivariance of the equivalence relation leads to
nonaxiomatizability of geometry. But we learn only axiomatizable geometries
in the last two thousand years. How can we guess, that nonaxiomatizable
geometries exist? The straight way from the Euclidean geometry to physical
geometries was very difficult, and the physical geometry has been obtained
on an oblique way.

J.L.Synge \cite{S60} has introduced the world function for
description of the Riemannian geometry. I did not know the papers
of Synge and introduced the world function for description of the
Riemannian space-time in general relativity. My approach differed
slightly from the approach of Synge. In particular, I had obtained
an equation for the world function of Riemannian geometry
\cite{R1962}.
\begin{equation}
\frac{\partial \sigma \left( x,x^{\prime }\right) }{\partial x^{i}}%
G^{ik^{\prime }}\frac{\partial \sigma \left( x,x^{\prime }\right) }{\partial
x^{\prime k}}=2\sigma \left( x,x^{\prime }\right) ,\qquad G^{ik^{\prime
}}G_{lk^{\prime }}=\delta _{l}^{i},\qquad G_{lk^{\prime }}\equiv \frac{%
\partial ^{2}\sigma \left( x,x^{\prime }\right) }{\partial x^{l}\partial
x^{\prime k}}  \label{a6.1}
\end{equation}%
This equation was obtained as a corollary of definition of the world
function of the Riemannian geometry as an integral along the geodesic,
connecting points $\ x$ and $x^{\prime }$. This equation contains only world
function and its derivatives.

This equation put the question. Let a world function does not satisfy the
equation. Does this world function describe a nonRiemannian geometry or it
does describe no geometry? It was very difficult to answer this question. On
one hand, the formalism, based on the world function, is a more developed
formalism, than formalism based on a use of metric tensor, because a
geodesic is described in terms of the world function by algebraic equation (%
\ref{c1.1}), whereas the same geodesic is described by differential
equations in terms the metric tensor.

On the other hand, the geodesic, described by (\ref{c1.1}) is
one-dimensional
only in the Riemannian geometry. In general, one equation (\ref{c1.1}) in $n$%
-dimensional space describes a $\left( n-1\right) $-dimensional surface. I
did not know, whether the surface is a generalization of a geodesic in any
geometry. I was not sure, because in the Euclidean geometry a straight
segment is one-dimensional be definition. I left this question unsolved and
returned to it almost thirty years later, in the beginning of ninetieth.

When the string theory of elementary particles appeared, it becomes clear
for me, that the particle may be described by means of a world surface
(tube) but not only by a world line. As the particle world line associates
with a geodesic, I decided, that a world tube may describe a particle. It
meant that there exist space-time geometries, where straights (geodesics)
are described by world tubes. The question on possibility of the physical
space-time geometry has been solved finally, when the quantum description
appeared to be a corollary of the space-time multivariance \cite{R91}.

\end{document}